\documentclass[a4paper,11pt]{article}
\usepackage{jcappub} % for details on the use of the package, please see the JCAP author manual
\usepackage{lineno}
\title{\boldmath Derivation of the electron and positron injection spectra from Geminga and Monogem}

\usepackage{orcidlink}
\author[a]{Qian Zhong\orcidlink{0009-0003-9058-1768}}

\affiliation[a]{MIFT Department, University of Messina, via F. S. D’Alcontres 31, I-98166 Messina, Italy}
\emailAdd{qian.zhong@studenti.unime.it}

\abstract{
Extended $\gamma$-ray emission has been observed around several nearby pulsars and is commonly interpreted as inverse-Compton radiation produced by relativistic electrons and positrons diffusing in the surrounding interstellar medium. In this work, a unified analysis of the halos associated with the Geminga and Monogem pulsars is presented, combining GeV--TeV $\gamma$-ray observations within a common physical framework.
Assuming continuous injection of $e^\pm$ pairs from the pulsar wind nebulae, the resulting $\gamma$-ray emission is modeled by accounting for particle diffusion and radiative energy losses. 
I find that the observed spectra of both Geminga and Monogem can be reproduced within this framework, provided that particle transport in the vicinity of the sources is significantly suppressed with respect to the average Galactic diffusion. The fits favor hard injection spectra and cutoff energies of order $10^5$--$10^6$~GeV, consistent with efficient lepton acceleration in pulsar environments.
Using the best-fit injection models inferred from the $\gamma$-ray data, then I estimate the contribution of Geminga and Monogem to the local cosmic-ray positron flux measured by AMS-02. I find that the slow-diffusion region surrounding the sources strongly suppresses the positron flux reaching the Earth, leading to a subdominant contribution over most of the AMS-02 energy range, with a possible effect only near the upper end of the measured spectrum.
The results support an interpretation in which TeV halos trace regions of inhibited particle diffusion around pulsars, while at the same time implying only a limited impact on the local positron flux. This combined analysis highlights the importance of extended $\gamma$-ray observations for constraining particle transport in the vicinity of Galactic cosmic-ray sources.
}

\begin{document}
\maketitle
\flushbottom

\section{Introduction}
\label{sec:intro}

The AMS-02 experiment has measured the cosmic-ray positron ($e^+$) flux with unprecedented precision from $\sim 0.5$~GeV up to $\sim 1$~TeV \cite{PhysRevLett.122.041102,AGUILAR2020}. Despite this progress, the physical origin of the positron spectrum --- and in particular the well-known positron excess at tens to hundreds of GeV --- remains debated.

At low energies, below $\sim 10$~GeV, the $e^+$ flux is expected to be dominated by secondary production, originating from inelastic interactions of primary cosmic rays with interstellar gas \cite{Delahaye:2008ua,Boudaud:2014dta,DiMauro:2014iia,Evoli:2020szd,Di_Mauro_2023}. Contemporary calculations indicate that secondaries can account for nearly the entire AMS-02 flux at $\sim 1$~GeV and for a substantial fraction (typically $50$--$70\%$) at $\sim 10$~GeV, depending on the assumed height of the diffusive halo \cite{Di_Mauro_2023}. At higher energies, however, a pronounced excess emerges: between $100$ and $1000$~GeV, the expected secondary contribution drops to the $\mathcal{O}(10\%)$ level relative to the data. This discrepancy cannot be removed by plausible uncertainties in propagation or in the relevant production cross sections, which are estimated to contribute only at the few-percent level to the overall uncertainty budget \cite{Orusa_2022}. These considerations strongly motivate the presence of one or more primary Galactic sources of high-energy positrons.

Rotation-powered pulsars and their pulsar wind nebulae (PWNe) are among the most compelling candidates \cite{Hooper:2008kg}. In these systems, a fraction of the pulsar spin-down power is converted into electron--positron pairs ($e^\pm$), which can be accelerated to very high energies \cite{Bykov:2017xpo,Amato:2020zfv}. Because $e^\pm$ suffer severe radiative losses through synchrotron emission and inverse Compton scattering during propagation, the observed positron spectrum above tens of GeV is expected to be dominated by relatively local sources, typically within $\lesssim 1$~kpc. This makes nearby middle-aged pulsars particularly relevant for interpreting AMS-02 data (see, e.g., \cite{hooperHAWCObservationsStrongly2017,Orusa:2021tts,Cholis:2022kio,Orusa:2024psrcat}).

A major development in the last decade has been the discovery of extended $\gamma$-ray emission around several pulsars --- often referred to as ``pulsar halos'' --- at GeV and TeV energies. TeV halos extending over several degrees have been detected by Milagro, HAWC, HESS, and LHAASO \cite{2009ApJ...700L.127A,Abeysekara:2017science,2023ApJ...944L..29A,HESS:2023sbf,2021PhRvL.126x1103A,cao2024}. At GeV energies, an extended halo around Geminga has been measured with \emph{Fermi}-LAT data \cite{DiMauro:2019yvh}. More generally, population studies and catalog-based searches suggest that halo-like emission may be common among Galactic pulsars \cite{Linden:2017vvb,DiMauro:2019hwn,DiMauro:2020jbz,Hooper:2021kyp}. These halos are naturally interpreted as inverse-Compton-scattering (ICS) emission from $e^\pm$ that escape the PWN and diffuse into the surrounding medium, upscattering photons of the interstellar radiation field (ISRF) to $\gamma$-ray energies \cite{Gaensler_2003,2017hsn..book.2159S}. While X-ray searches for extended emission around Geminga have not yet yielded a clear detection, they have placed constraints suggesting a relatively weak ambient magnetic field, likely of order $\lesssim \mu{\rm G}$ \cite{Liu:2019sfl,Khokhriakova:2023rqh,Manconi:2024wlq}.

Beyond establishing the presence of high-energy leptons, $\gamma$-ray halos provide a direct probe of particle transport near pulsars. The multi-degree extent of the TeV halos around Geminga (PSR~J0633+1746) and Monogem (PSR~B0656+14) implies diffusion coefficients suppressed by $\sim 2$--$3$ orders of magnitude relative to the canonical Galactic value inferred from cosmic-ray nuclei data \cite{Abeysekara:2017science,Korsmeier:2021bkw,trottaCONSTRAINTSCOSMICRAYPROPAGATION2011,berezinskiiv.s.AstrophysicsCosmicRays1990}. This inhibited-diffusion picture poses a significant theoretical challenge and has crucial implications for pulsar interpretations of the positron excess, since strong confinement near the source can severely limit the fraction of injected $e^\pm$ that reach the Solar neighborhood \cite{Evoli:2018aza,Lopez-Coto:2017pbk,Liu:2019zyj,Fang:2019iym,Recchia:2021kty,PhysRevD.105.123008}. Recent HAWC observations have substantially improved the precision of the TeV-halo spectral energy distributions and surface-brightness profiles around Geminga and Monogem, providing correspondingly tighter constraints on the underlying lepton injection and transport parameters \cite{HAWC:2024GemingaMonogem}. For a recent review of $\gamma$-ray halos around pulsars, see \cite{Amato:2024dss}.

From a phenomenological standpoint, pulsar-based models have long been shown to reproduce AMS-02 positron data when including contributions from a small number of nearby objects and/or the cumulative emission from the Galactic pulsar population \cite{Mlyshev:2009twa,Boudaud:2014dta,2014JCAP...04..006D,Manconi:2016byt,Fornieri_2020,DiMauro:2020cbn,Cholis_2018,Evoli_2021,Manconi:2020ipm,Orusa_2021}. Many studies find that explaining the positron excess typically requires converting $\sim 1$--$10\%$ of the pulsar spin-down energy into escaping $e^\pm$ pairs. In the previous work \cite{Orusa_2021}, they performed realistic simulations of the Galactic pulsar population calibrated to AMS-02 $e^+$ data, finding that the smoothness of the measured spectrum suggests a dominant contribution from a few nearby and middle-aged pulsars. An important next step is therefore to exploit multi-wavelength observations to constrain the injection and transport properties of the most relevant nearby sources.

In this paper, I focus on Geminga and Monogem, two benchmark nearby pulsars with well-measured $\gamma$-ray halos at GeV and TeV energies. I combine the GeV halo constraints derived with \emph{Fermi}-LAT \cite{DiMauro:2019yvh} with the latest HAWC measurements of the TeV halos \cite{HAWC:2024GemingaMonogem} to infer the lepton injection properties and the energy dependence of diffusion in the vicinity of the sources. I then compute the resulting $e^\pm$ flux at Earth and assess the possible contribution of Geminga and Monogem to the positron excess. By jointly exploiting GeV and TeV halo observations within a unified transport framework, I aim to construct a self-consistent picture of lepton escape, propagation, and radiative signatures from these nearby systems.

The paper is organized as follows. In Section~\ref{sec:model}, the physical model for $e^\pm$ injection, transport, and the associated $\gamma$-ray emission is introduced. In Section~\ref{sec:results}, the constraints from the combined GeV--TeV halo data set and the implications for the positron flux at Earth are presented. Finally, in Section~\ref{sec:conclusions}, the main findings are summarized and their broader impact on pulsar interpretations of the positron excess is discussed.

\section{Physical model}
\label{sec:model}

The extended $\gamma$-ray emission around Geminga and Monogem is modeled as ICS emission produced by relativistic electrons and positrons ($e^\pm$) injected by the PWN and propagating in the surrounding interstellar medium (ISM).
This framework is widely used in the interpretation of pulsar halos from GeV to TeV energies and enables a direct connection between the $\gamma$-ray morphology and spectrum and the underlying transport of $e^\pm$ pairs. In particular, I aim at a unified description of the TeV halos measured by HAWC \cite{HAWC:2024GemingaMonogem} and the GeV halo constraints derived from \emph{Fermi}-LAT data \cite{DiMauro:2019yvh}, spanning $\gamma$-ray energies from a few GeV to beyond $10^2$~TeV.

\subsection{Source properties: Geminga and Monogem}
\label{subsec:pulsars}

The pulsar parameters entering the modeling are taken from the ATNF catalog and coincide with those adopted in the latest HAWC halo analysis \cite{HAWC:2024GemingaMonogem}. Geminga (PSR~J0633+1746) is a nearby, middle-aged pulsar at a distance of $\sim 250$~pc, with characteristic age $\sim 370$~kyr and present-day spin-down luminosity $\dot E \simeq 3.2 \times 10^{34}$~erg~s$^{-1}$. It is radio-quiet and is one of the best-studied local pulsars across the electromagnetic spectrum, with a well-established extended $\gamma$-ray halo detected both at GeV energies with \emph{Fermi}-LAT and at TeV energies with HAWC. Monogem (PSR~B0656+14) is similarly nearby ($\sim 288$~pc), but younger ($\sim 110$~kyr) and slightly more powerful ($\dot E \simeq 3.8 \times 10^{34}$~erg~s$^{-1}$); unlike Geminga, it is radio-loud and is also associated with a prominent TeV halo. Their proximity and ages make them prime candidates for contributing to the local $e^\pm$ spectrum, while the measured GeV/TeV halo morphologies provide direct constraints on the diffusion and energy-loss history of the injected leptons. I summarize the key source properties in Table~\ref{tab:pulsarprops}.

For nearby sources, the time delay $d/c$ is negligible at the level of the modeling. The source age is therefore set to $T \simeq t_{\rm obs}$.When needed, one may instead define $T = t_{\rm obs} + d/c$, as in \cite{HAWC:2024GemingaMonogem}; this difference is numerically irrelevant here.

\begin{table}[t]
\centering
\caption{Reference parameters for Geminga (PSR~J0633+1746) and Monogem (PSR~B0656+14) used in this work, consistent with ATNF and with the latest HAWC analysis \cite{HAWC:2024GemingaMonogem}.}
\label{tab:pulsarprops}
\begin{tabular}{lcc}
\hline\hline
 & Geminga (J0633+1746) & Monogem (B0656+14) \\
\hline
Distance $d$ & $250$ pc & $288$ pc \\
Observed age $t_{\rm obs}$ & $370$ kyr & $110$ kyr \\
Spin-down luminosity $\dot E$ & $3.2 \times 10^{34}$ erg s$^{-1}$ & $3.8 \times 10^{34}$ erg s$^{-1}$ \\
\hline\hline
\end{tabular}
\end{table}

\subsection{Injection of electrons and positrons}
\label{subsec:injection}

It is assumed that the PWN continuously injects $e^\pm$ pairs into the ISM with a spectrum
\begin{equation}
Q(E,t) \;=\; Q_0(t)\,
\left(\frac{E}{E_0}\right)^{-\gamma_e}\,
\exp\!\left(-\frac{E}{E_c}\right),
\label{eq:inj}
\end{equation}
where $E_0 = 1$~GeV is a reference energy, $\gamma_e$ is the injection index, and $E_c$ is the exponential cutoff. The TeV-halo spectra measured by HAWC require $E_c$ to be at least several tens of TeV, and the absence of a sharp cutoff in the observed $\gamma$-ray spectrum typically implies leptons extending to energies of at least a few $10^2$~TeV \cite{HAWC:2024GemingaMonogem,Abeysekara2017}. In other words, the highest-energy HAWC photons require very energetic parent electrons and positrons, even if they do not necessarily imply injection all the way to the PeV scale in every modeling setup.

The time dependence of the injected power is tied to the pulsar spin-down luminosity. Assuming magnetic-dipole braking with braking index $k = 3$, the spin-down luminosity evolves as
\begin{equation}
\dot E(t) \;=\; \dot E_0 \left(1+\frac{t}{\tau_0}\right)^{-2},
\label{eq:edot_t}
\end{equation}
where $\tau_0$ is the characteristic spin-down timescale. In TeV-halo studies, one often adopts $\tau_0 \simeq 12$~kyr, following the HAWC convention \cite{HAWC:2024GemingaMonogem}. The $e^\pm$ luminosity is parameterized as
\begin{equation}
L_{e^\pm}(t) \;=\; \eta\,\dot E(t),
\label{eq:lept_lum}
\end{equation}
where $\eta$ is the $e^\pm$ conversion efficiency. The normalization $Q_0(t)$ in Eq.~\eqref{eq:inj} is fixed by requiring that the injected energy rate equals $L_{e^\pm}(t)$ above a minimum energy $E_{\min}$:
\begin{equation}
\int_{E_{\min}}^\infty dE\,E\,Q(E,t) \;=\; L_{e^\pm}(t).
\label{eq:qnorm}
\end{equation}

The total rotational energy released over the pulsar lifetime up to age $T$ is
\begin{equation}
W_0(T) \;=\; \int_0^{T} dt\,\dot E(t)
\;=\; \dot E(T)\,T\left(1+\frac{T}{\tau_0}\right),
\label{eq:w0_exact}
\end{equation}
which, in the commonly used limit $T \gg \tau_0$, reduces approximately to
\begin{equation}
W_0(T) \;\simeq\; \tau_0\,\dot E(T)\left(1+\frac{T}{\tau_0}\right)^2.
\label{eq:w0_approx}
\end{equation}
The total energy injected into pairs is then $E_{\rm tot} = \eta W_0$. The latest HAWC analysis finds efficiencies of order a few percent for Geminga and Monogem, consistent with the level typically required in pulsar interpretations of the positron excess \cite{HAWC:2024GemingaMonogem,DiMauro:2019yvh,Orusa:2024psrcat}.

\subsection{Transport: diffusion--loss equation and diffusion models}
\label{subsec:transport}

The propagation of $e^\pm$ is described by the diffusion--loss equation
\begin{equation}
\frac{\partial \mathcal{N}_e(E,\mathbf{r},t)}{\partial t}
\;=\;
\nabla\!\cdot\!\left[D(E,\mathbf{r})\,\nabla \mathcal{N}_e\right]
\;+\;
\frac{\partial}{\partial E}\!\left[b(E)\,\mathcal{N}_e\right]
\;+\;
Q(E,t)\,\delta^{(3)}\!\left(\mathbf{r}-\mathbf{r}_s(t)\right),
\label{eq:diff_loss}
\end{equation}
where $\mathcal{N}_e(E,\mathbf{r},t)$ is the differential number density per unit energy, $D(E,\mathbf{r})$ is the diffusion coefficient, $b(E)\equiv -dE/dt$ is the energy-loss rate, and $\mathbf{r}_s(t)$ is the source position.

At energies above a few GeV, energy losses are dominated by inverse Compton scattering and synchrotron radiation:
\begin{equation}
b(E) \;=\; b_{\rm IC}(E)+b_{\rm syn}(E)
\;\simeq\; b_0\,E^2,
\label{eq:losses}
\end{equation}
where $b_0$ depends on the local magnetic field and radiation energy density. When needed, Klein--Nishina effects can be included directly in the $\gamma$-ray emissivity (see below), while retaining the quadratic approximation for transport as a useful baseline description.

Two transport setups commonly used to interpret pulsar halos are considered.

\smallskip
\noindent{\it (i) One-zone (uniform) diffusion:}
\begin{equation}
D(E) \;=\; D_0\left(\frac{E}{1\,{\rm GeV}}\right)^\delta,
\label{eq:diff_1zone}
\end{equation}
with $\delta$ typically fixed to a Kolmogorov-like value, $\delta = 1/3$.

\smallskip
\noindent{\it (ii) Two-zone diffusion (slow zone + ISM):}
motivated by the fact that TeV-halo data imply strongly suppressed diffusion only within tens to hundreds of parsecs around the pulsar, while the large-scale ISM must retain a much larger diffusion coefficient to reproduce cosmic-ray nuclei data. I adopt
\begin{equation}
D(E,r) \;=\;
\begin{cases}
D_{\rm in}(E) = D_{{\rm in},0}\left(\dfrac{E}{1\,{\rm GeV}}\right)^\delta, & r<r_b,\\[6pt]
D_{\rm out}(E)= D_{{\rm out},0}\left(\dfrac{E}{1\,{\rm GeV}}\right)^\delta, & r\ge r_b,
\end{cases}
\label{eq:diff_2zone}
\end{equation}
where $r_b$ is the radius of the inhibited-diffusion region. This formulation is particularly relevant when connecting the halo morphology to the positron flux at Earth: HAWC constrains diffusion at TeV-relevant energies to be much smaller than Galactic-average values (e.g.\ $D$ at $E_e \sim 100$~TeV of order a few $\times 10^{27}$~cm$^2$~s$^{-1}$ around Geminga and Monogem) \cite{HAWC:2024GemingaMonogem}, while \emph{Fermi}-LAT halo constraints probe the $\sim 10^2$~GeV lepton regime \cite{DiMauro:2019yvh}.

For Geminga, the pulsar proper motion can affect the GeV halo morphology because GeV-emitting electrons have longer cooling times and can travel farther before losing energy. Following Ref.~\cite{DiMauro:2019yvh}, this effect is implemented by allowing a time-dependent source position $\mathbf{r}_s(t)$, approximated as rectilinear motion in the plane of the sky.

\subsection{Solution for the $e^\pm$ density}
\label{subsec:solution}

The transport equation in Eq.~\eqref{eq:diff_loss} admits an analytic Green-function solution in the case of homogeneous diffusion or, more generally, within each region of piecewise homogeneous transport, with appropriate matching at the boundary. For continuous injection from a source located at $\mathbf{r}_s(t')$, the lepton density at time $T$ can be written as a convolution of the injection term with the propagator (e.g., \cite{Syrovatskii1959,Atoyan:1995gj,Yuksel:2008zj}).

A convenient parametrization is obtained by introducing the injection energy $E_s$ of particles that are observed with energy $E$ after a time interval $\Delta t$:
\begin{equation}
\Delta t \;=\; \int_E^{E_s}\frac{dE'}{b(E')}\,,
\qquad \Rightarrow \qquad
E_s(E,\Delta t)=\frac{E}{1-b_0E\Delta t}
\qquad {\rm for}\qquad b(E)=b_0E^2,
\label{eq:Es}
\end{equation}
and the associated diffusion--loss length
\begin{equation}
\lambda^2(E,E_s) \;=\; 4\int_E^{E_s} dE'\,\frac{D(E')}{b(E')}\,.
\label{eq:lambda}
\end{equation}
In terms of these quantities, the lepton density at the observation time $T$ reads
\begin{equation}
\mathcal{N}_e(E,\mathbf{r},T)
=
\int_0^{T} dt'\,
\frac{b\!\left(E_s\right)}{b(E)}\,
\frac{1}{\left(\pi\lambda^2\right)^{3/2}}
\exp\!\left[-\frac{\left|\mathbf{r}-\mathbf{r}_s(t')\right|^2}{\lambda^2}\right]\,
Q\!\left(E_s,t'\right),
\label{eq:Ne_cont}
\end{equation}
where $E_s \equiv E_s(E,T-t')$ and $\lambda^2 \equiv \lambda^2(E,E_s)$. Equation~\eqref{eq:Ne_cont} is the central quantity entering both the ICS $\gamma$-ray emission and the local $e^\pm$ flux at Earth.

HAWC halo observations indicate that diffusion can be strongly suppressed in the vicinity of some pulsars, while other measurements favor a much larger diffusion coefficient in the surrounding ISM \cite{Abeysekara:2017science,Hooper:2017gtd}. A widely used phenomenological approach to reconcile these facts is a two-zone transport model, in which slow diffusion operates inside a bubble of radius $r_b$ around the source, while standard diffusion applies outside \cite{Fang:2018fjy,Tang:2018wyr}. In its simplest form one adopts
\begin{equation}
D(r,E)=
\begin{cases}
D_{\rm in}\left(\dfrac{E}{1~{\rm GeV}}\right)^{\delta}, & 0<r<r_b,\\[6pt]
D_{\rm out}\left(\dfrac{E}{1~{\rm GeV}}\right)^{\delta}, & r\ge r_b,
\end{cases}
\label{eq:D_2zone_piecewise}
\end{equation}
with the same energy-scaling index $\delta$ in both regions. Physical mechanisms that could generate such an effective suppression --- for example, self-generated turbulence driven by steep cosmic-ray gradients --- have been discussed in \cite{Evoli:2018swy}.

For a sharp transition at $r=r_b$, the solution for continuous injection can still be written in the form of Eq.~\eqref{eq:Ne_cont}, but with a modified spatial kernel that enforces continuity of the diffusive flux across the boundary. Following \cite{Tang:2018wyr} (see also \cite{Fang:2018fjy}), one can express the density as
\begin{equation}
\mathcal{N}_e(E,\mathbf{r},T)
=
\int_0^{T} dt'\,
\frac{b(E_s)}{b(E)}\,
Q(E_s,t')\,
\mathcal{H}\!\left(\Delta r;E,E_s\right),
\label{eq:Ne_cont_2zone}
\end{equation}
where $\Delta r \equiv |\mathbf{r}-\mathbf{r}_s(t')|$, and $\mathcal{H}$ depends on the two diffusion lengths
\[
\lambda_{\rm in}^2=4\int_E^{E_s} dE'\,\frac{D_{\rm in}(E')}{b(E')},
\qquad
\lambda_{\rm out}^2=4\int_E^{E_s} dE'\,\frac{D_{\rm out}(E')}{b(E')}.
\]
Defining $\xi \equiv \sqrt{D_{\rm in}/D_{\rm out}}$ and $\epsilon \equiv r_b/\lambda_{\rm in}$, the kernel can be written as
\begin{align}
\mathcal{H}(\Delta r;E,E_s) &=
\frac{\xi(\xi+1)}{(\pi \lambda_{\rm in}^2)^{3/2}\,
\Big[2\xi^2\,{\rm erf}(\epsilon)-\xi(\xi-1)\,{\rm erf}(2\epsilon)+2\,{\rm erfc}(\epsilon)\Big]}
\nonumber\\
&\times
\begin{cases}
\exp\!\left(-\dfrac{\Delta r^2}{\lambda_{\rm in}^2}\right)
+\left(\dfrac{\xi-1}{\xi+1}\right)\left(\dfrac{2r_b}{r}-1\right)
\exp\!\left[-\dfrac{(\Delta r-2r_b)^2}{\lambda_{\rm in}^2}\right],
& r<r_b,\\[10pt]
\left(\dfrac{2\xi}{\xi+1}\right)\left[\dfrac{r_b}{r}+\xi\left(1-\dfrac{r_b}{r}\right)\right]
\exp\!\left\{-\left[\dfrac{\Delta r-r_b}{\lambda_{\rm out}}+\dfrac{r_b}{\lambda_{\rm in}}\right]^2\right\},
& r\ge r_b,
\end{cases}
\label{eq:H_kernel_2zone}
\end{align}
where $r \equiv |\mathbf{r}|$. In the limit $D_{\rm in} = D_{\rm out}$, or for $r_b \gg \Delta r$, Eqs.~\eqref{eq:Ne_cont_2zone}--\eqref{eq:H_kernel_2zone} reduce to the homogeneous solution in Eq.~\eqref{eq:Ne_cont}, as expected.

The two-zone model introduces additional degrees of freedom ($D_{\rm in}$, $D_{\rm out}$, and $r_b$), and these can be partially degenerate, especially for very extended halos where different parameter combinations can produce similar surface-brightness profiles. Moreover, the transition at $r_b$ may in reality be smooth rather than discontinuous.

To keep the $\gamma$-ray analysis tractable, the minimal one-zone setup is adopted for the spectral fits, and the $e^+$ flux at Earth is subsequently computed within a representative two-zone framework. This procedure captures the leading impact of transport uncertainties on the local positron flux while avoiding an underconstrained multi-parameter fit to the halo morphology. For the spectral energy distributions considered here, a one-zone description is a reasonable approximation because most of the line-of-sight emissivity comes from the slow-diffusion region close to the source. A full fit to the spatial profile would, however, require the explicit two-zone treatment.

In the positron-flux calculation presented below, a halo size of $r_b = 100$~pc is adopted. This choice is motivated by the angular extension of the Geminga halo inferred from \emph{Fermi}-LAT data in \cite{dimauroDetectionGrayHalo2019}: an angular scale of about $20^\circ$--$30^\circ$ at a distance of $\sim 250$~pc corresponds to a physical size of order $10^2$~pc.

\subsection{Inverse-Compton $\gamma$-ray emission}
\label{subsec:ic_gamma}

The differential IC emissivity, namely the number of photons emitted per unit volume, time, and energy, is
\begin{equation}
j_{\rm IC}(E_\gamma,\mathbf{r})
=
\int_{E_{\min}}^\infty dE\,
\mathcal{N}_e(E,\mathbf{r},T)\,
\mathcal{P}_{\rm IC}(E,E_\gamma),
\label{eq:jic}
\end{equation}
where $\mathcal{P}_{\rm IC}(E,E_\gamma)$ is the IC power emitted into photons of energy $E_\gamma$ by an electron or positron of energy $E$. Using the full Klein--Nishina cross section \cite{BlumenthalGould},
\begin{equation}
\mathcal{P}_{\rm IC}(E,E_\gamma)
=
c\,E_\gamma
\int d\epsilon\, n_{\rm ph}(\epsilon)\,
\frac{d\sigma_{\rm IC}}{dE_\gamma}(E,\epsilon,E_\gamma),
\label{eq:pic_general}
\end{equation}
with $n_{\rm ph}(\epsilon)$ the target photon density of the ISRF. The ISRF is modeled as the sum of CMB, IR, and optical/starlight components, adopting the local radiation-field model commonly used in halo analyses~\cite{Strong:1998fr,Strong:2000tw}.

The observable differential $\gamma$-ray intensity (flux per solid angle) at an angular separation $\theta$ from the pulsar is obtained by integrating along the line of sight:
\begin{equation}
\frac{d\Phi_\gamma}{dE_\gamma\,d\Omega}(E_\gamma,\theta)
=
\frac{1}{4\pi}
\int_0^\infty ds\, j_{\rm IC}\!\left(E_\gamma,\mathbf{r}(s,\theta)\right),
\label{eq:gamma_intensity}
\end{equation}
where $\mathbf{r}(s,\theta)$ is the position along the line of sight. The integrated flux in a region $\Delta\Omega$ is then
\begin{equation}
\frac{d\Phi_\gamma}{dE_\gamma}(E_\gamma,\Delta\Omega)
=
\int_{\Delta\Omega} d\Omega\,
\frac{d\Phi_\gamma}{dE_\gamma\,d\Omega}(E_\gamma,\theta).
\label{eq:gamma_flux_region}
\end{equation}
Equations~\eqref{eq:jic}--\eqref{eq:gamma_flux_region} directly connect the transport model to the HAWC-measured surface-brightness profiles and spectra \cite{HAWC:2024GemingaMonogem} and to the \emph{Fermi}-LAT GeV halo constraints \cite{DiMauro:2019yvh}.

\subsection{Positron flux at Earth}
\label{subsec:positrons_earth}

The interstellar positron flux at Earth is computed from the local positron density:
\begin{equation}
\Phi_{e^+}^{\rm IS}(E)
=
\frac{c}{4\pi}\,
\mathcal{N}_{e^+}(E,\mathbf{r}_\oplus,T),
\label{eq:phi_is}
\end{equation}
where $\mathbf{r}_\oplus$ is the Earth position. For symmetric pair injection one has $\mathcal{N}_{e^+} = \mathcal{N}_e/2$.

When comparing to AMS-02 below $\sim 10$~GeV, solar modulation may be included through the force-field approximation:
\begin{equation}
\Phi_{e^+}^{\rm TOA}(E_{\rm TOA})
=
\left(\frac{E_{\rm TOA}^2-m_e^2}{E_{\rm IS}^2-m_e^2}\right)\,
\Phi_{e^+}^{\rm IS}(E_{\rm IS}),
\qquad
E_{\rm IS}=E_{\rm TOA}+|Z|e\,\phi,
\label{eq:ffa}
\end{equation}
where $\phi$ is the Fisk potential and $Z=+1$ for positrons.

TeV-halo measurements constrain $D(E)$ at electron energies of tens to hundreds of TeV through the morphology at multi-TeV $\gamma$-ray energies \cite{HAWC:2024GemingaMonogem}. GeV halo measurements probe the $\sim 10^2$~GeV lepton regime and are therefore directly relevant to the energy range of the positron excess \cite{DiMauro:2019yvh}. By combining the GeV and TeV halo information, one can constrain the injection parameters $(\gamma_e,E_c,\eta)$ and the characteristic slow-diffusion scale, and then derive the corresponding prediction for $\Phi_{e^+}(E)$ at Earth.

\section{Results}
\label{sec:results}

The goal of this analysis is to constrain the lepton injection spectra of the nearby PWNe Geminga and Monogem. I do so by fitting the spectral energy distributions (SEDs) of their extended $\gamma$-ray halos measured at GeV energies with \emph{Fermi}-LAT \cite{dimauroDetectionGrayHalo2019} and at TeV energies with HAWC \cite{HAWC:2024GemingaMonogem}. Once the injection parameters are determined from the $\gamma$-ray data, the corresponding $e^\pm$ population is propagated to the Solar system and the associated positron flux is computed in order to assess the possible contribution of these sources to the positron excess measured by AMS-02~\cite{AMS:2021nhj}.

\subsection{Fit to the GeV--TeV halo spectra}

The extended emission is modeled within the continuous-injection framework described in Section~\ref{sec:model}. In the fitting procedure, the $e^\pm$ conversion efficiency $\eta$ (which sets the overall normalization), the injection spectral index $\gamma_e$ (which controls the spectral slope), and the cutoff energy $E_c$ are varied. The diffusion-coefficient normalization is fixed to $D_0 = 10^{26}\,\mathrm{cm^2\,s^{-1}}$, and $\delta = 1/3$ is adopted, which are representative values for the inhibited diffusion inferred in halo studies~\cite{dimauroDetectionGrayHalo2019,HAWC:2019tcx,DiMauro:2019hwn,HAWC:2024GemingaMonogem}. In this setup, $D_0$ primarily controls the spatial extension of the halo, whereas the broadband SED is driven mainly by the injection history and radiative losses within the region of interest. $D_0$ is therefore kept fixed in this spectral-focused analysis.

\begin{figure}[t]
    \centering
    \includegraphics[width=0.49\textwidth]{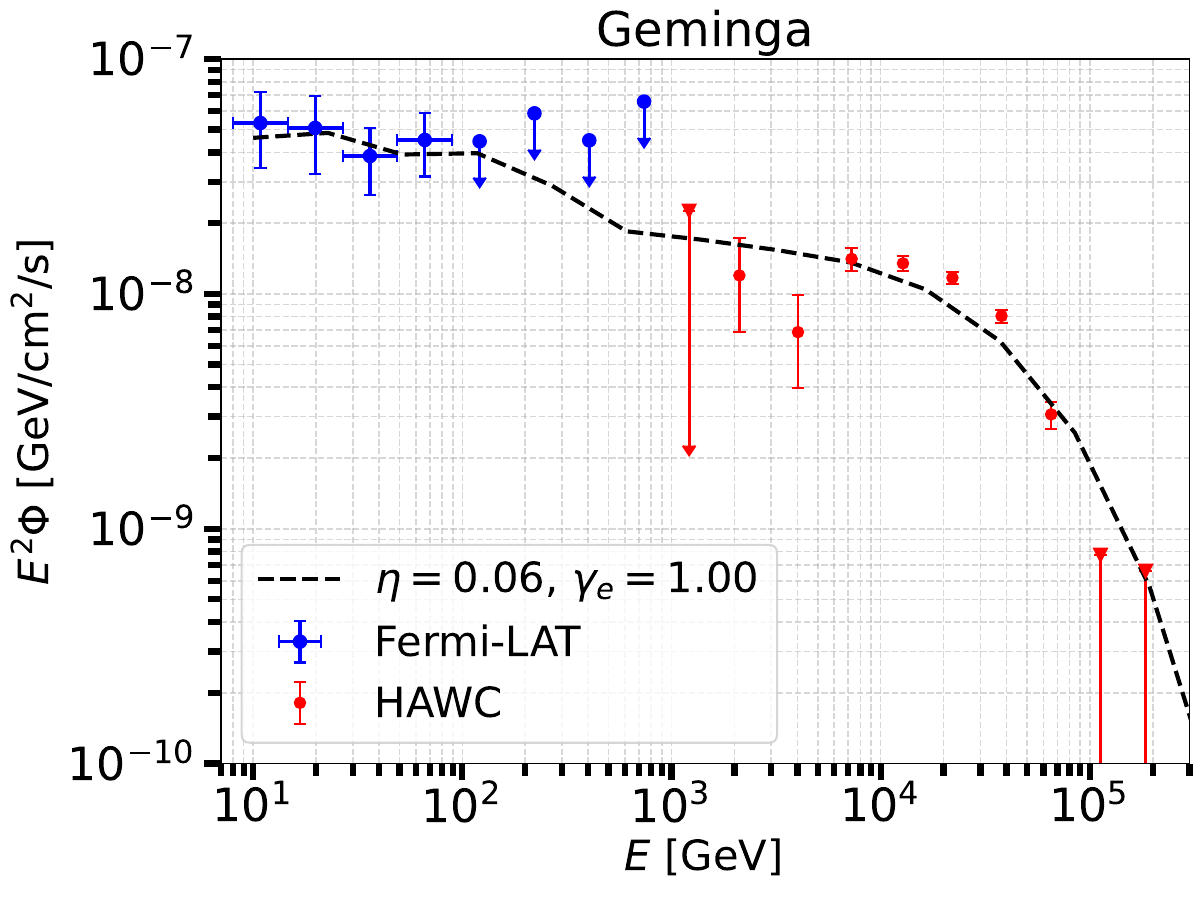}
    \includegraphics[width=0.49\textwidth]{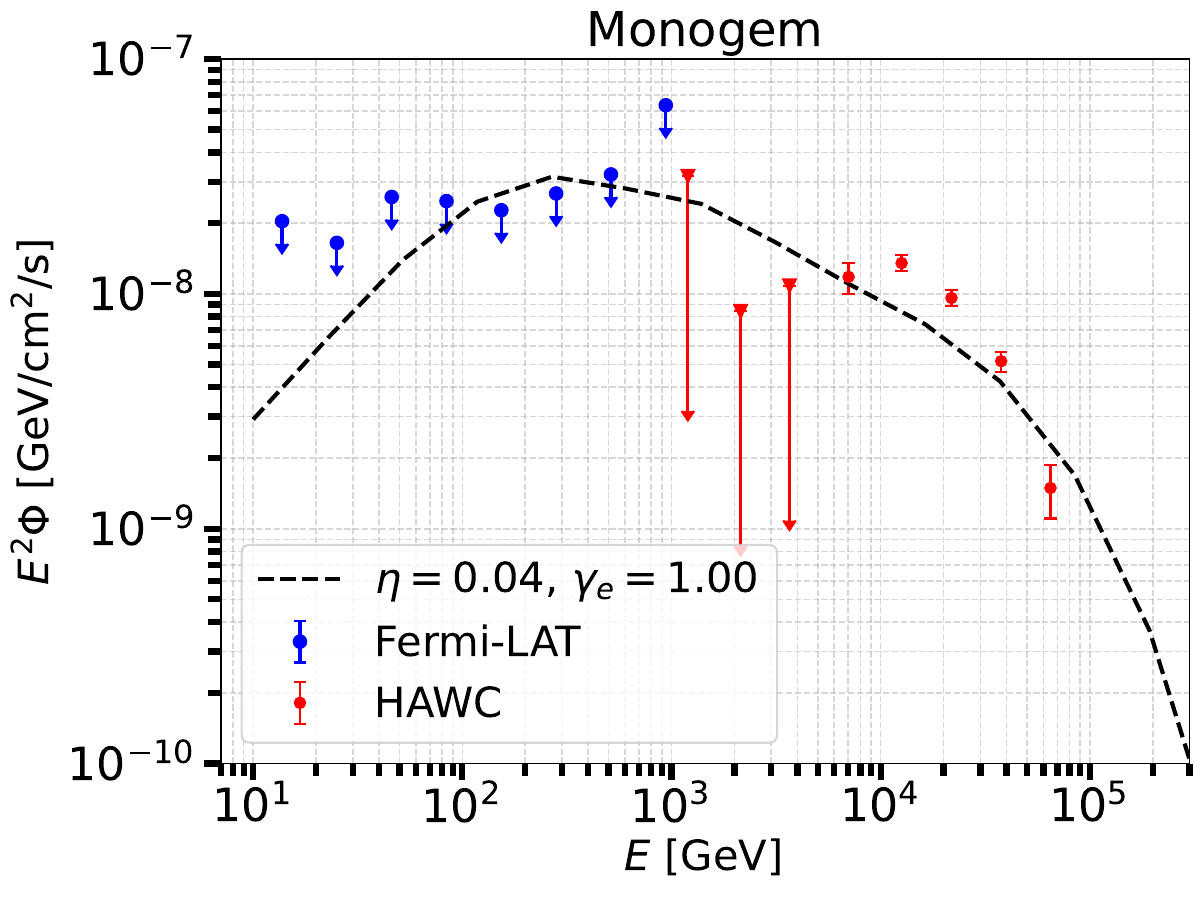}
    \caption{Best-fit $\gamma$-ray spectral energy distributions of the Geminga (left) and Monogem (right) halos in the continuous-injection scenario. The black dashed lines show the model predictions, while the blue and red points correspond to the \emph{Fermi}-LAT \cite{dimauroDetectionGrayHalo2019} and HAWC \cite{HAWC:2024GemingaMonogem} data, respectively. For the $\gamma$-ray calculation, a one-zone transport model is adopted with a diffusion coefficient $D = 10^{26}\,\mathrm{cm^2\,s^{-1}}$ at 1~GeV and $\delta = 1/3$.}
    \label{fig:gemmon_gamma}
\end{figure}

Figure~\ref{fig:gemmon_gamma} shows that the model provides a good joint description of the GeV \emph{Fermi}-LAT points and the multi-TeV HAWC measurements, reproducing the smooth transition between the two energy regimes for both sources. A large cutoff energy is essential to sustain inverse-Compton emission up to the highest energies observed by HAWC. With a significantly smaller $E_c$, the model would not contain enough very-high-energy leptons to generate $\gamma$ rays above $\sim 10$~TeV.

The best-fit parameters are very similar for the two sources, with a hard injected spectrum, $\gamma_e \simeq 1.0$, and a high-energy cutoff $E_c \simeq 10^5\,\mathrm{GeV}$. The required efficiencies are $\eta \simeq 0.06$ for Geminga and $\eta \simeq 0.04$ for Monogem. These values are comparable to those obtained in \cite{dimauroDetectionGrayHalo2019} and are broadly consistent with the efficiencies inferred by the HAWC Collaboration in physically motivated ICS-based analyses \cite{HAWC:2024GemingaMonogem}.

The very hard injection indices preferred by the updated fit differ from those typically inferred before the publication of the new HAWC data set. For example, \cite{dimauroDetectionGrayHalo2019}, using the earlier HAWC measurements \cite{HAWC:2019tcx}, found that a significantly softer injection index, of order $\gamma_e \sim 1.8$--$1.9$, was sufficient to reproduce the $\gamma$-ray data. The difference is readily understood: the updated HAWC fluxes are systematically higher than the earlier ones at multi-TeV energies, and therefore require a harder injected lepton spectrum to maintain enough power at the highest energies.

\begin{figure}[t]
    \centering
    \includegraphics[width=.49\textwidth]{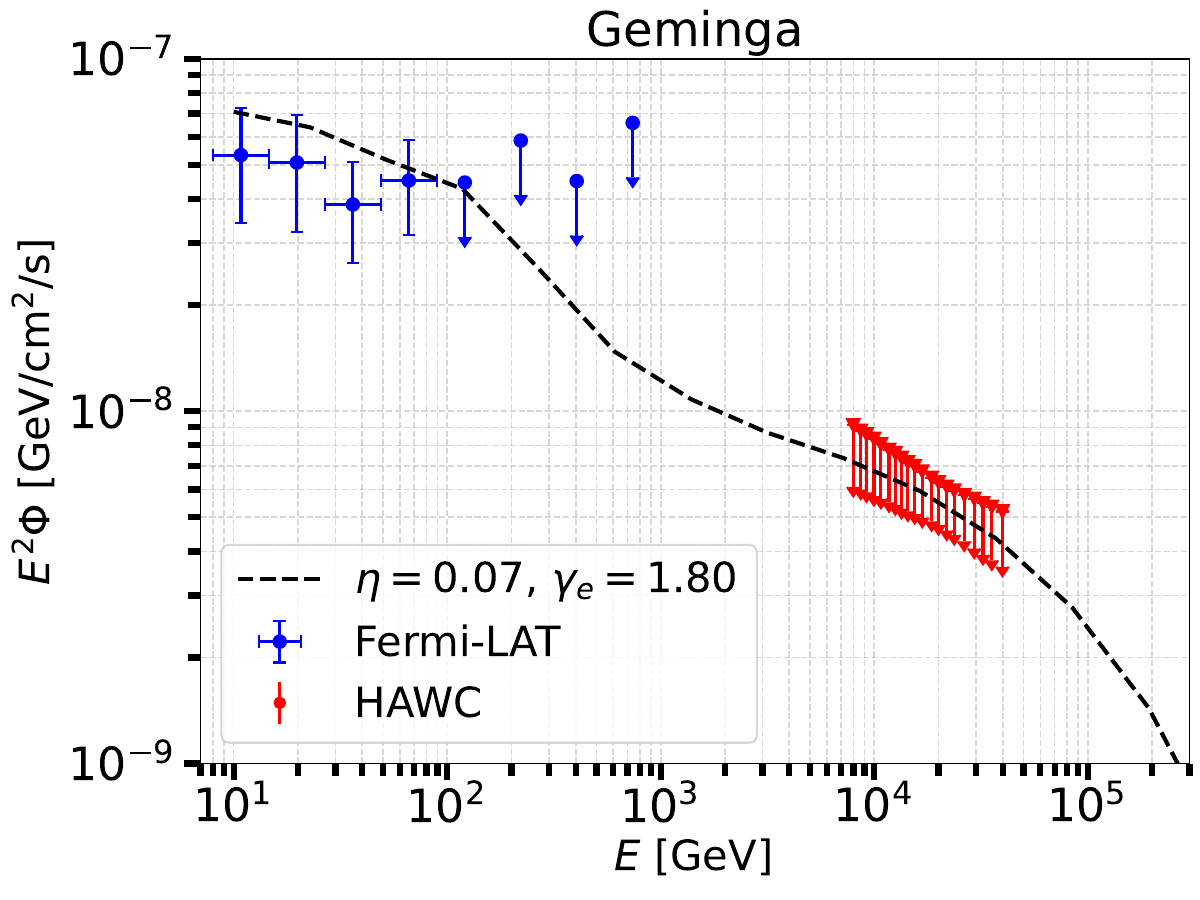}
    \includegraphics[width=.49\textwidth]{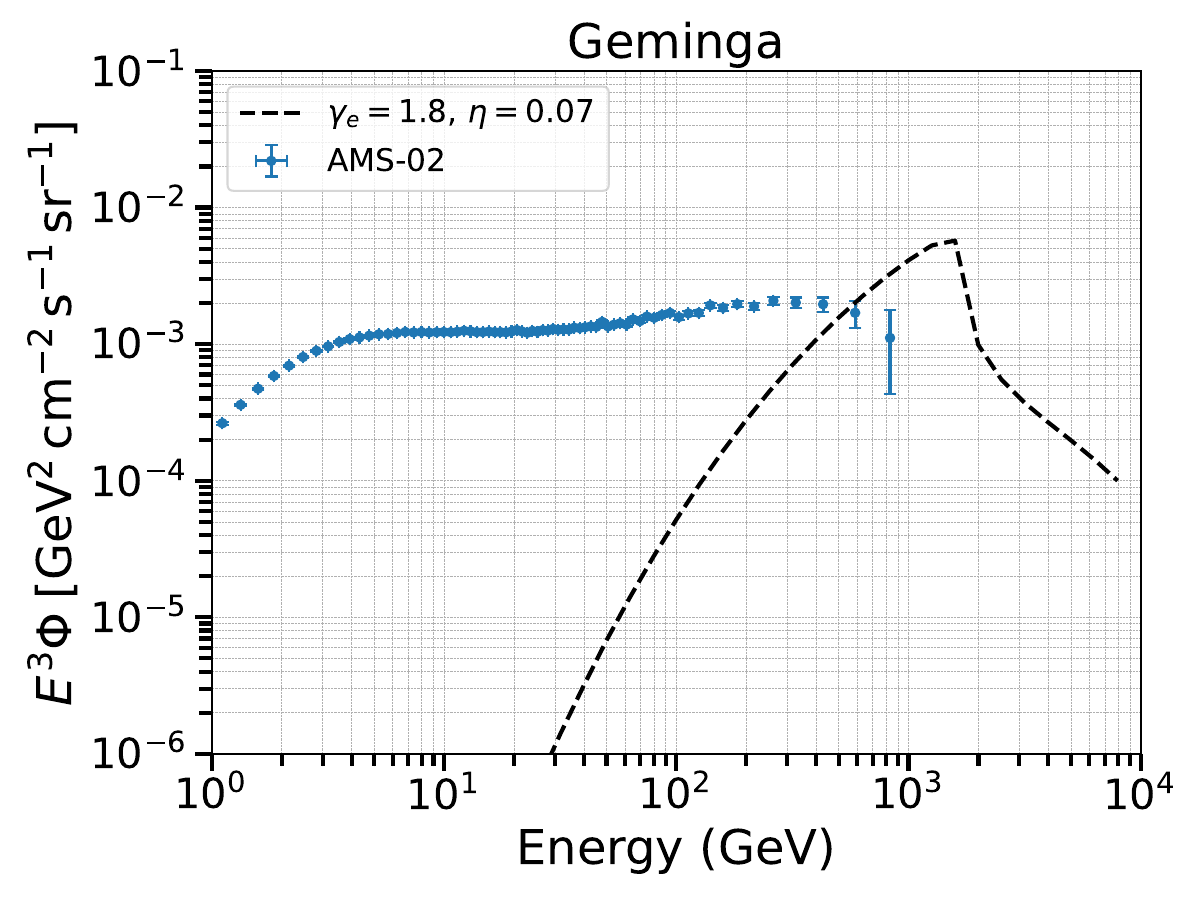}
    \caption{Comparison fit for Geminga obtained using the older HAWC data set. The left panel shows the best-fit $\gamma$-ray spectral energy distribution, $E^2 \Phi_\gamma$, obtained by fitting the \emph{Fermi}-LAT data \cite{DiMauro:2019yvh} together with the earlier HAWC measurement \cite{Abeysekara:2017science}, for $E_c = 10^6$~GeV and $\gamma_e = 1.8$. The right panel shows the corresponding positron energy flux at Earth, $E^3 \Phi_{e^+}$, for the same injection model.}
    \label{fig:gemingasofter}
\end{figure}

For comparison, the fit is repeated using the older HAWC data set~\cite{Abeysekara:2017science}. In that case, a much softer injection slope provides an acceptable joint fit to the \emph{Fermi}-LAT and HAWC data. This is illustrated in Figure~\ref{fig:gemingasofter}, where $\gamma_e = 1.8$ and $\eta \simeq 0.07$. The figure makes clear that, when the older HAWC flux normalization is adopted as in \cite{DiMauro:2019yvh}, the inferred injection spectrum is softer and the corresponding positron contribution at Earth is substantially larger. This comparison helps explain why earlier studies could assign a more important role to Geminga in the interpretation of the AMS-02 positron excess.

\subsection{Predicted positron flux at Earth}

\begin{figure}[t]
    \centering
    \includegraphics[width=.49\textwidth]{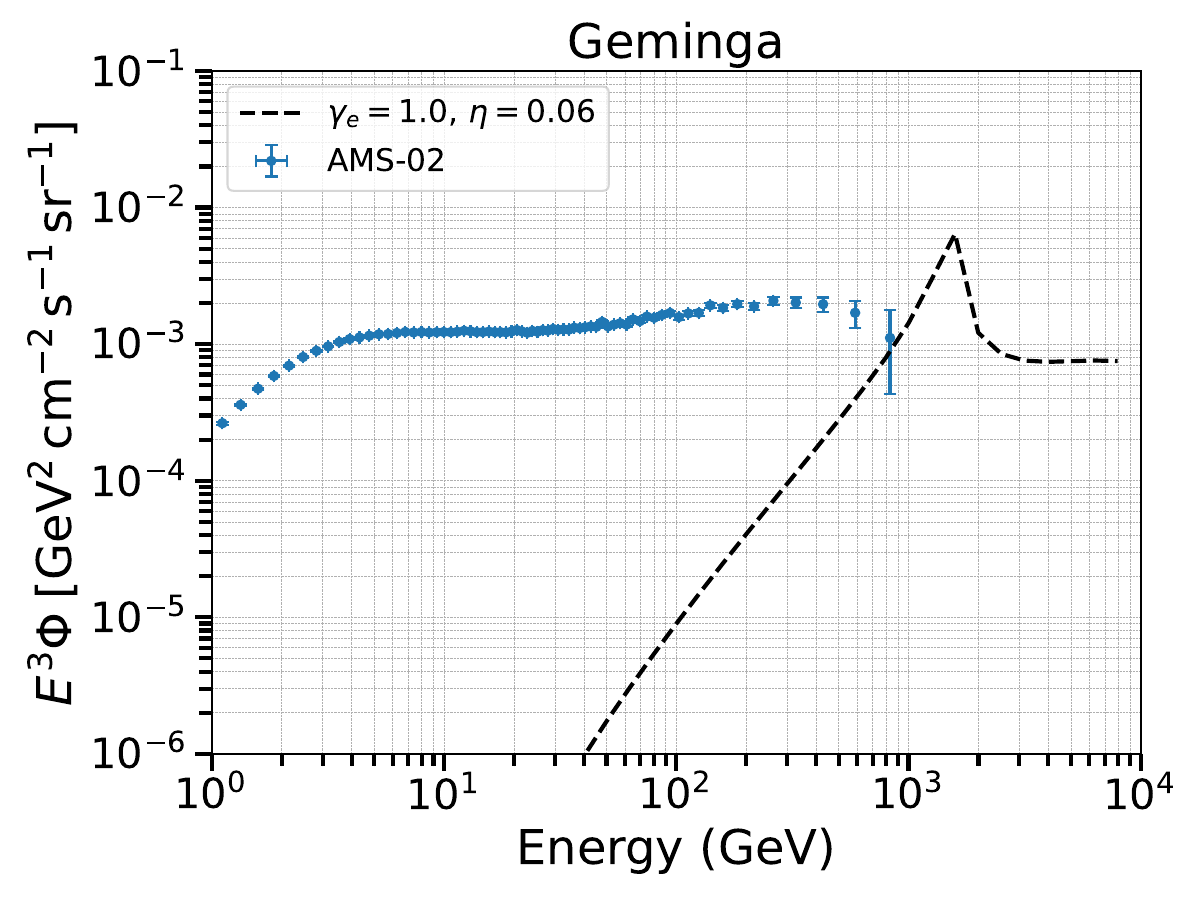}
    \includegraphics[width=.49\textwidth]{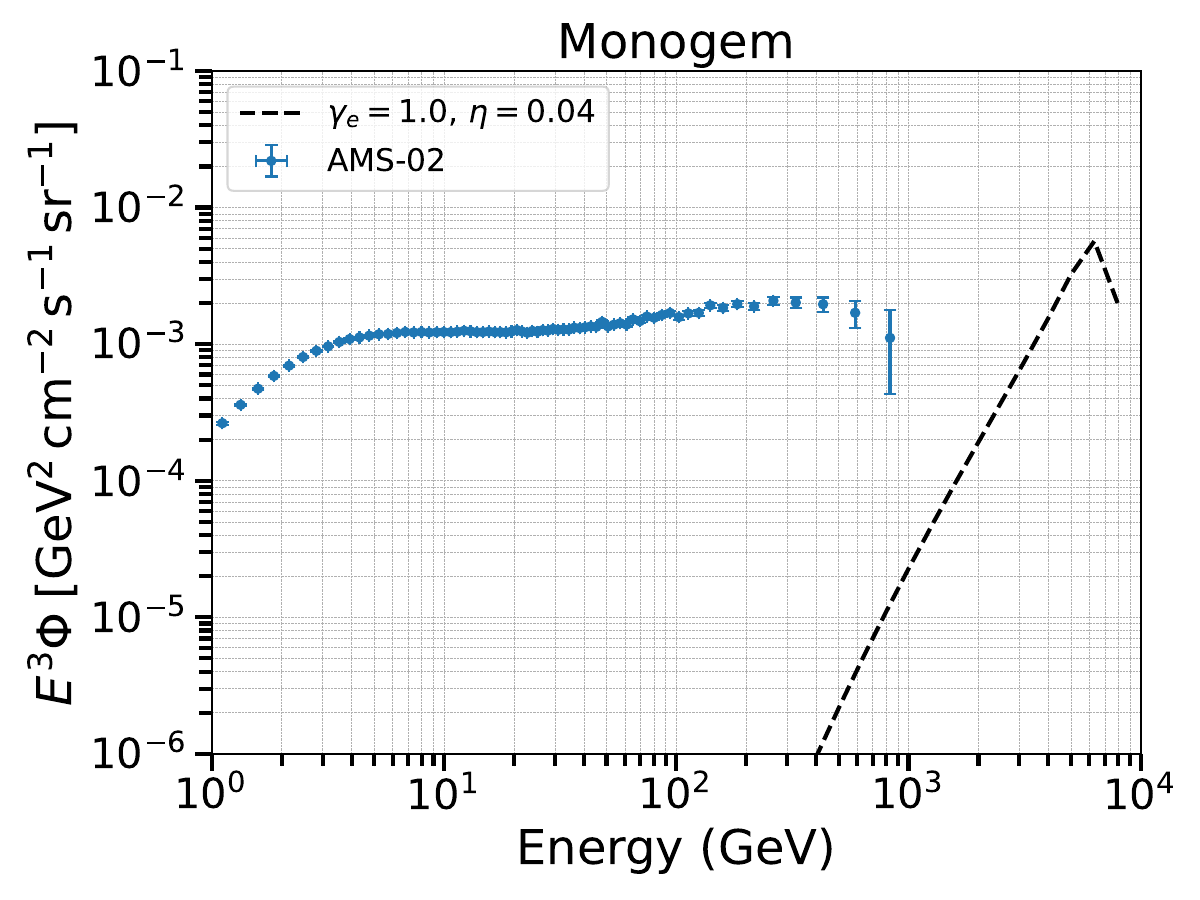}
    \caption{Predicted positron energy flux at Earth, $E^3 \Phi_{e^+}$, from Geminga (left) and Monogem (right) for the best-fit injection parameters shown in Figure~\ref{fig:gemmon_gamma}, compared with AMS-02 measurements. For the positron-flux calculation a two-zone model is adopted with $r_b = 100$~pc, $D_{\rm in} = 10^{26}\,\mathrm{cm^2\,s^{-1}}$, and $D_{\rm out} = 10^{28}\,\mathrm{cm^2\,s^{-1}}$ at 1~GeV, with the same energy dependence in both regions.}
    \label{fig:gemmon_ams}
\end{figure}

Using the same best-fit injection parameters, the positron flux at Earth is computed and compared with the AMS-02 data. The predicted source contributions are shown in Figure~\ref{fig:gemmon_ams}. For the inhibited-diffusion setup adopted here, the propagated positron flux from both sources is strongly suppressed throughout the energy range in which AMS-02 measures the bulk of the excess, namely $\sim 10$--$500$~GeV. In other words, the small diffusion coefficient favored by the halo emission implies that only a limited fraction of the injected leptons can escape efficiently enough to reach the Solar neighborhood before suffering substantial radiative losses.

At higher energies, the predicted spectra become much harder and rise toward the TeV scale. This behavior reflects the interplay between transport and losses: in the presence of strong confinement near the source, only the highest-energy leptons can contribute efficiently at Earth, leading to a hard high-energy tail in the local spectrum. However, even in this regime the source contribution remains below the observed AMS-02 positron flux, and the very hard spectral shape implied by the updated $\gamma$-ray fit is not naturally aligned with the apparent softening of the AMS-02 spectrum above a few hundred GeV.

Overall, the results show that the continuous-injection model, with parameters that successfully reproduce the GeV--TeV $\gamma$-ray halos of Geminga and Monogem, implies only a limited contribution to the local positron flux over most of the AMS-02 range. This conclusion differs from those of earlier studies such as \cite{dimauroDetectionGrayHalo2019,hooperHAWCObservationsStrongly2017}. The origin of the difference is straightforward: those analyses relied on the older HAWC data set, whose lower flux normalization admitted a softer injection spectrum, and therefore a much larger positron contribution at Earth.

Any sizable impact of Geminga or Monogem on the local positron spectrum would occur, if at all, mainly in the highest-energy tail, where current measurements are statistically limited and where additional nearby sources and/or different transport conditions may become important.

\section{Conclusions and discussion}
\label{sec:conclusions}

In this work, the extended $\gamma$-ray halos surrounding the Geminga and Monogem PWNe have been modeled under the assumption of continuous injection of relativistic electron--positron pairs, spatial diffusion, and energy losses dominated by inverse Compton scattering and synchrotron radiation.

A single class of injection and transport models provides a satisfactory description of the $\gamma$-ray emission from both sources. In particular, the model naturally accounts for the smooth spectral connection between the GeV measurements from \emph{Fermi}-LAT \cite{DiMauro:2019yvh} and the multi-TeV data from HAWC \cite{HAWC:2024GemingaMonogem}, while requiring only moderate lepton-conversion efficiencies to match the observed luminosities. The updated HAWC data favor hard injection spectra, with $\gamma_e \simeq 1$, and cutoff energies of order $10^5$~GeV or larger.

Using the same best-fit injection parameters, and adopting a representative two-zone transport setup for propagation to the Earth, I find that the positron flux from both pulsars is strongly suppressed in the energy range probed by AMS-02, namely $\sim 10$--$500$~GeV. This suppression arises from the inefficient escape implied by the small diffusion coefficient within a slow-diffusion region of characteristic size of order $10^2$~pc. As a result, Geminga and Monogem contribute only a subdominant fraction of the measured positron spectrum below the TeV domain. The predicted positron spectra exhibit a characteristic hard rise toward higher energies, reflecting the fact that only the highest-energy particles can reach the Solar system before suffering severe radiative losses, but even this feature remains below the AMS-02 measurements.

These findings reinforce the emerging picture that diffusion coefficients on the order of $10^{26}\,\mathrm{cm^2\,s^{-1}}$ are typical within the $\sim 100$~pc halos of middle-aged pulsars. At the same time, they show that bright, extended TeV $\gamma$-ray emission can coexist with a negligible impact on the local cosmic-ray positron flux when transport near the source is sufficiently slow.

It is important to stress, however, that the conclusions regarding the positron contribution at Earth are derived within a simplified transport prescription. In the present work the $\gamma$-ray SEDs are fitted in a one-zone setup, while the positron flux is computed in a representative two-zone model with a sharp transition at $r_b$. More realistic scenarios --- including a smooth spatial transition, anisotropic transport, or time-dependent diffusion --- could increase the efficiency of particle escape at large distances and therefore modify the predicted positron flux. A dedicated study of these effects is left for future work.

In summary, within the framework of continuous injection and the transport assumptions adopted here, Geminga and Monogem provide a consistent explanation of their observed extended $\gamma$-ray halos while playing only a marginal role in the cosmic-ray positron excess measured near Earth. Explaining the bulk of the excess therefore likely requires additional nearby pulsars, a broader Galactic pulsar population, or a different source component.

\acknowledgments
Acknowledgments will be added in the final version.

\bibliographystyle{apsrev4-1}
\bibliography{main}

\end{document}